\documentclass{article}

\usepackage{arxiv}

\usepackage[utf8]{inputenc} 
\usepackage[T1]{fontenc}    
\usepackage{hyperref}       
\usepackage{url}            
\usepackage{booktabs}       
\usepackage{amsfonts}       
\usepackage{nicefrac}       
\usepackage{microtype}      
\usepackage{lipsum}		
\usepackage{graphicx}
\usepackage{natbib}
\usepackage{doi}
\usepackage{amsmath}

\title{Dataset of Fluorescence Spectra and Chemical Parameters of Olive Oils}

\author{ \href{https://orcid.org/0000-0003-2562-9932}{\includegraphics[scale=0.06]{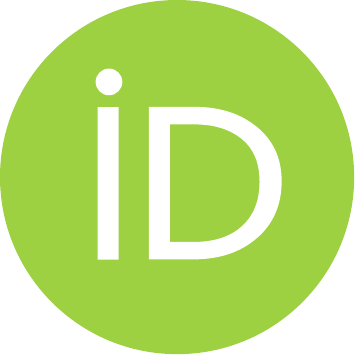}\hspace{1mm}Francesca Venturini }\thanks{Contact email: vent@zhaw.ch} \\
	Institute of Applied Mathematics and Physics\\ 
 Zurich University of Applied Sciences\\Winterthur, Switzerland, \texttt{vent@zhaw.ch} \\Artificial Intelligence Research and Development\\
 TOELT LLC, Switzerland \\
	\And
	\href{https://orcid.org/0000-0000-0000-0000}{\includegraphics[scale=0.06]{orcid.pdf}\hspace{1mm}Michela Sperti} \\
	PolitoBIOMed Lab\\Department of Mechanical and\\Aerospace Engineering\\
 Politecnico di Torino, Turin, Italy
 \And
	\href{https://orcid.org/0000-0000-0000-0000}{\includegraphics[scale=0.06]{orcid.pdf}\hspace{1mm}Umberto Michelucci} \\
	Artificial Intelligence Research and Development\\ TOELT LLC, Switzerland\\
	\texttt{umberto.michelucci@toelt.ai} \\
 Computer Science Department \\
 Lucerne  University of Applied Sciences and Arts\\ Lucerne, Switzerland\\
 \And
	\href{https://orcid.org/0000-0000-0000-0000}{\includegraphics[scale=0.06]{orcid.pdf}\hspace{1mm}Arnaud Gucciardi} \\
	Artificial Intelligence Research and Development\\ TOELT LLC, Switzerland\\
	\texttt{arnaud.gucciardi@toelt.ai} \\
 Artificial Intelligence Laboratory\\University of Ljubljana, Ljubljana, Slovenia
 \And
	\href{https://orcid.org/0000-0000-0000-0000}{\includegraphics[scale=0.06]{orcid.pdf}\hspace{1mm}Vanessa M. Martos} \\
	Department of Plant Physiology\\
 Faculty of Sciences\\Biotechnology Institute\\University of Granada, Spain
 \And
	\href{https://orcid.org/0000-0000-0000-0000}{\includegraphics[scale=0.06]{orcid.pdf}\hspace{1mm}Marco A. Deriu} \\
	PolitoBIOMed Lab\\Department of Mechanical and\\Aerospace Engineering\\
 Politecnico di Torino, Turin, Italy
}

\renewcommand{\shorttitle}{Dataset of Fluorescence Spectra and Chemical Parameters of Olive Oils}

\hypersetup{
pdftitle={Dataset of Fluorescence Spectra and Chemical Parameters of Olive Oils},
pdfsubject={},
pdfauthor={Francesca Venturini},
pdfkeywords={Fluorescence; Olive Oil; Chemical Parameters; Quality control},
}

\begin{document}
\maketitle

\begin{abstract}
	This dataset encompasses fluorescence spectra and chemical parameters of 24 olive oil samples from the 2019–2020 harvest provided by the producer Conde de Benalúa, Granada, Spain. The oils are characterized by different qualities: 10 extra virgin olive oil (EVOO), 8 virgin olive oil (VOO), and 6 lampante olive oil (LOO) samples. For each sample, the dataset includes fluorescence spectra obtained with two excitation wavelengths, oil quality, and five chemical parameters necessary for the quality assessment of olive oil. The fluorescence spectra were obtained by exciting the samples at 365 nm and 395 nm under identical conditions. The dataset includes the values of the following chemical parameters for each olive oil sample: acidity, peroxide value, $K_{270}$, $K_{232}$, ethyl esters, and the quality of the samples (EVOO, VOO, or LOO). The dataset offers a unique possibility for researchers in food technology to develop machine learning models based on fluorescence data for the quality assessment of olive oil due to the availability of both spectroscopic and chemical data. The dataset can be used, for example, to predict one or multiple chemical parameters or to classify samples based on their quality from fluorescence spectra.
\end{abstract}

\keywords{Fluorescence \and Olive Oil \and Chemical Parameters \and Quality control}

\section{Summary}


The dataset presented is a compilation of measurements of analytical chemistry and fluorescence spectroscopy. The dataset includes fluorescence spectra and chemical parameters of 24 Spanish olive oils from the 2019–2020 harvest. The 24 samples were collected at SCA San Sebastián Puente del Ventorro, Benalua de las Villas, Spain.
The data were later measured at the Institute of Applied Mathematics and Physics, Zurich University of Applied Sciences, Technikumstrasse 9, 8401 Winterthur, Switzerland. The fluorescence spectroscopy data was acquired by a miniature spectrometer with a 1024 element CCD array that acquires the entire spectrum in one single measurement. The dataset includes a total of 960 spectra (24 oil samples × 2 excitation wavelengths x 20 repeated measurements). Each of the 960 spectra is an array of 1024 values whose elements are the intensity at the different pixel positions. 
The chemical parameters were determined by accredited laboratories using the procedures described in the European Commission regulation and its amendment  \cite{european2013commission,regulation1991commission}. These regulations control the methods for the quality assessment of olive oils and provide a decision tree to verify whether an olive oil class is consistent with the declared quality.

\noindent The value of the dataset for research purposes is summarized in the points below.
\begin{itemize}
    \item The data are useful for studying the link between optical properties (fluorescence and absorption spectroscopy), chemical characteristics (such as oil acidity, peroxide value, and fatty acid content), and olive oil quality  (extra virgin, virgin, and lampante olive oil).
    \item This dataset is the first available that contains fluorescence spectra and  chemical analysis obtained by accredited laboratories on samples coming from a single producer.
    \item Many researchers can benefit from the data: computer scientists can use the data to develop machine learning models that link optical to chemical properties; researchers in food technology that are interested in studying chemical properties of olive oil samples of different qualities; engineers that want to develop new optical analysis techniques alternative to the current expensive and time-consuming analytical chemistry methods.
    \item This dataset can be used to perform explainability analysis to identify spectral characteristics that are related to different chemical properties (e.g., the acidity of the oil). An example is given in the paper \cite{venturini2023extraction}. This will further advance the understanding of the complex chemical composition of olive oil and its link to its quality and health benefits.
    \item This dataset can be used to develop instruments based on fluorescence spectroscopy for the rapid and cost-effective quality assessment of olive oil.

\end{itemize}

\section{Data Description}

The dataset consists of one CSV file that contains the columns described in Table \ref{table:features}.
\begin{table}[th!]
\begin{tabular}{p{2.5cm}cp{32em}}
\specialrule{.2em}{0em}{0em} 
\textbf{Feature}          & \textbf{Datatype} & \textbf{Description}      \\[0.1cm]
\specialrule{.1em}{0em}{0em}  
Sample & String & Oil sample name: the values are  'D03','D04','D05', 'D06', 'D07' ,'D08', 'D09', 'D10', 'D
19', 'D20', 'D35', 'D38', 'D45', 'D46', 'D47', 'D49', 'D51', 'D52', 'D53', 'D64', 'D77', 'D81', 'D92','D73'\\ 
Repetition & Integer & Repetition number. There are 20 repetition for each oil and led: the iteration number goes from 0 to 19)\\ 
Led & Integer & Excitation LED identifier: 1 (395 nm), 2 (365 nm)\\ 
Data & Float & The fluorescence spectra. The feature is a string composed of 1024 values given between square brackets and seprated by a comma, as for example [1491.0, 1508.0, ..., 1545.0]. Each value is the raw intensity of the fluorescence signal at the given pixel of the detector of the spectrometer.\\ 
Quality & String & Quality of the oil. Possible values are ‘EXTRA’, ‘VIRGIN’, ‘LAMPANTE’\\ 
FAEES & Float & Fatty acid ethyl esters in mg/Kg: content of waxes, fatty acid methyl esters and fatty acid ethyl esters\\
K232 & Float & UV Absorbance at 232 nm ($K_{270}$)\\ 
K270 & Float & UV Absorbance at 270 nm ($K_{232})$\\ 
Acidity & Float & Acidity: expressed as percentage (\%) of oleic acid \\ 
Peroxide Index & Float & Quantity of those substances in the sample, expressed in terms of milliequivalents of active oxygen per kilogram (mEqO2/Kg), which oxidize potassium iodide.\\[0.1cm] \specialrule{.2em}{0em}{0em} \smallskip
\end{tabular}
\caption{Information on each feature available in the dataset.}
\label{table:features}
\end{table}

A background file\footnote{Fluorescence\_olive\_oil\_dataset\_background.csv} is also provided. The file contains 1024 values that correspond to the intensity measured by the spectrometer without any light (dark counts). This spectrum can be subtracted from the raw fluorescence spectra to remove the effect of the dark counts. The same file can be used for the spectra taken at both 365 nm and 395 nm.

\begin{figure}[hbt]
    \centering
    \includegraphics[width=13cm]{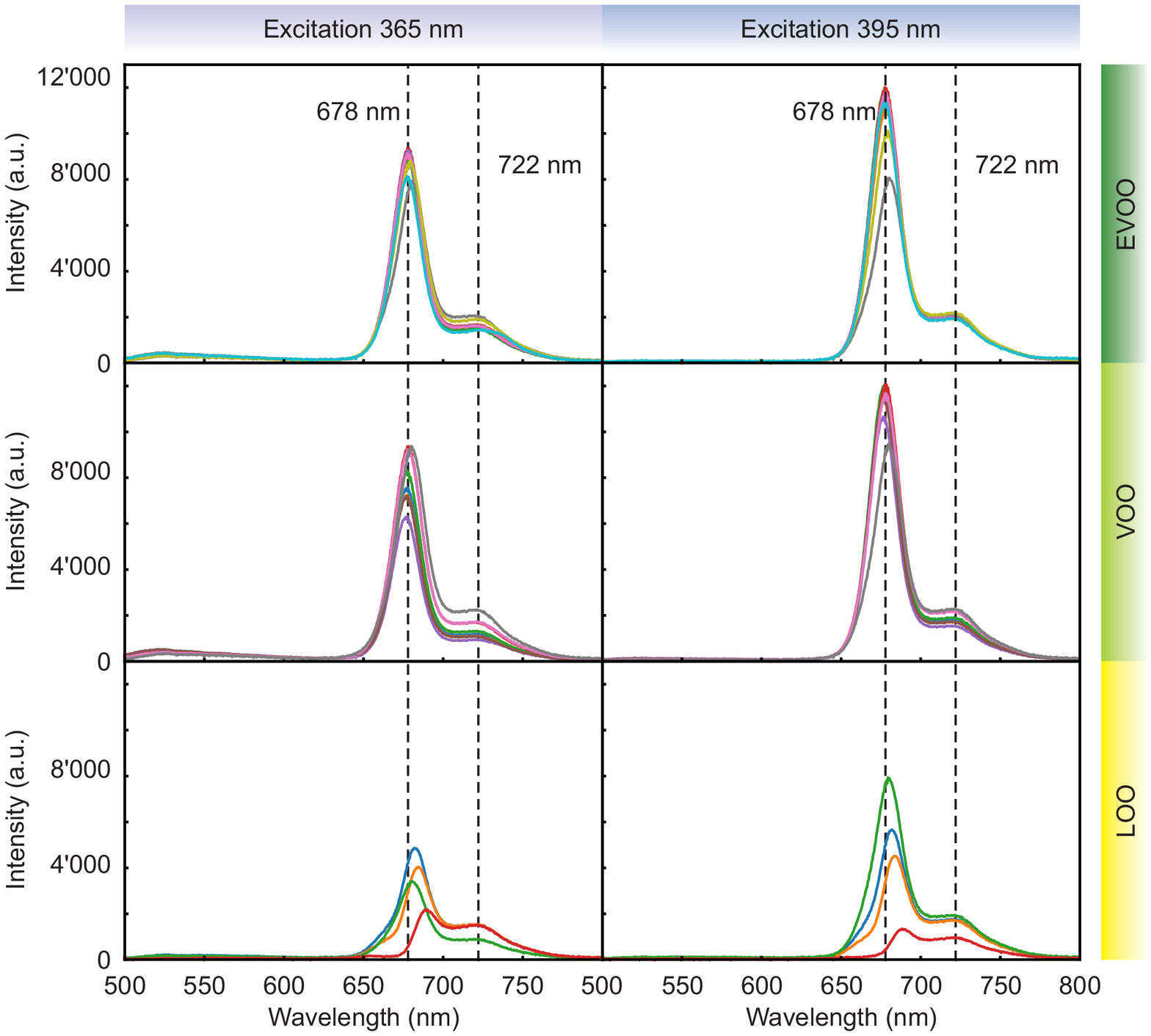}
    \caption{Fluorescence emission spectra of selected olive oils divided in the quality classes EVOO, VOO and LOO. On the left: spectra obtained with excitation at 365 nm; on the right: spectra obtained with excitation at 395 nm. Each curve shows a single spectrum without averaging or smoothing after the background subtraction. Reproduced from \cite{venturini2023extraction}.}
    \label{fig:spectra}
\end{figure}

The raw fluorescence spectra of selected oils obtained with excitation at 365 nm and 395 nm are shown in Fig. \ref{fig:spectra}.

\section{Materials and methods}

\subsection{Olive Oil Samples}

The dataset contains the fluorescence spectra and the chemical parameters of 24 oils. The oils are characterized by different quality categories: 10 extra virgin olive oil (EVOO), 8 virgin olive oil (VOO), and 6 lampante olive oil (LOO) samples. All samples were provided by Conde de Benalúa, Granada, southern Spain, and were prepared from the 2019–2020 harvest.
The properties and values of the chemical parameters of the oil samples are listed in Table \ref{tab:oils}. 

\begin{table*}[hbt]
\centering
\begin{tabular}{lcccccr} 
\specialrule{.2em}{0em}{0em} 
Label	& Acidity & Peroxide value & $K_{270}$ & $K_{232}$  & FAEES & Quality \\
	&  (\%) & (mEq O$_2$/kg) &   & & (mg/Kg) &  \\
\specialrule{.1em}{0em}{0em}  
         D03 & 0.35 & 8.4 & 0.123 & 1.435 & 26 & VOO\\
         D04 & 0.34 & 8.6 & 0.108 & 1.403  & 40 & VOO\\
         D05 & 0.36 & 10.3 & 0.112 & 1.44  & 18 & VOO\\    
         D06 & 0.31 & 9.2 & 0.151 & 1.484 & 18 & VOO\\    
         D07 & 0.50 & 8.9 & 0.150 & 1.537 & 47 & VOO\\
         D08 & 0.40 & 8.5 & 0.158 & 1.546  & 25 & VOO\\
         D09 & - & - & - & - &  - & LOO\\
         D10 & - & - & - & - &  - & LOO\\
         D19 & 0.25 & 4.9 & 0.13 & 1.540  & 10 & EVOO\\
         D20 & 0.26 & 4.6 & 0.14 & 1.540 & 10 & EVOO\\
         D35 & 0.17 & 6.4 & 0.12 & 1.63  & 8 & EVOO\\
         D38 & 0.16 & 6.4 & 0.12 & 1.63  & 9 & EVOO\\
         D45 & 0.17 & 4.9 & 0.12 & 1.63  & 7 & EVOO\\
         D46 & 0.18 & 5.0 & 0.13 & 1.63  & 8 & EVOO\\
         D47 & 0.18 & 5.2 & 0.13 & 1.64  & 16 & EVOO\\
         D49 & 0.9 & 9.9 & - & - &  - & LOO\\
         D51 & 2.16 & - & - & - & - &LOO\\
         D52 & 1.78 & 22 & - & - & - & LOO\\
         D53 & 0.7 & 8.7 & - & - &  - & LOO\\
         D64 & 0.2 & 7.1 & 0.13 & 1.63  & 29 & VOO\\
         D73 & 0.2 & 8.9 & 0.14 & 1.66  & 15 & EVOO\\
         D77 & 0.24 & 10.4 & 0.13 & 1.74 & 26 & VOO\\
         D81 & 0.16 & 4.9 & 0.12 & 1.63  & 9 & EVOO\\
         D92 & 0.18 & 5 & 0.17 & 1.91 & 15 & EVOO\\
\specialrule{.2em}{0em}{0em} 
\end{tabular}
\caption{List of olive oil samples and their physicochemical characteristics. FAEES: fatty acid ethyl esters, EVOO: extra virgin olive oil, VOO: virgin olive oil, LOO: lampante olive oil.\label{tab:oils}}
\end{table*}

For data acquisition, the samples were placed in commercial 4 ml clear glass vials, taking care that no headspace was present to reduce oxidation. All oils were stored in the dark and at 20 °C during the entire time of the measurements. 

\subsection{Fluorescence Data Acquisition}

\begin{figure}[hbt]
    \centering
    \includegraphics[width=4.5cm]{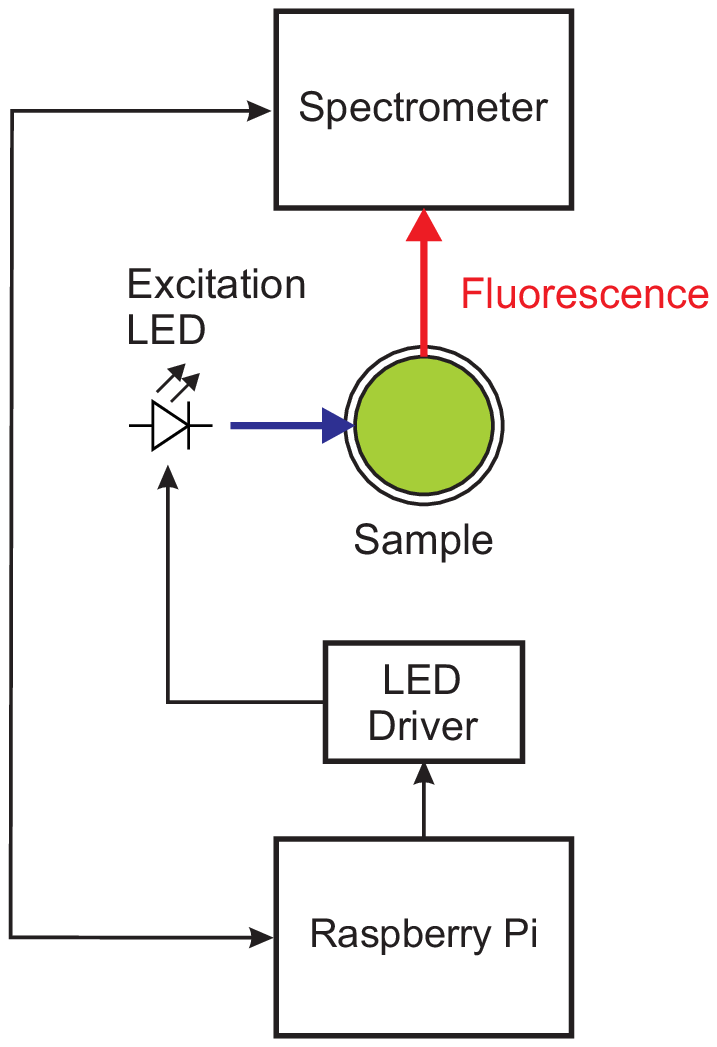}
    \caption{Schematics of the portable fluorescence sensor. Blue: excitation light, red: fluorescence light. From \cite{venturini2021exploration}.}
    \label{fig:schematics}
\end{figure}

The fluorescence spectroscopy data were acquired using the portable sensor described in \cite{venturini2021exploration}. Since already published, only the most relevant characteristics are reported here. The reader is referred to this publication for more details. The schematic design of the spectrometer is sown in Fig. \ref{fig:schematics}. The excitation light was provided by two UV LEDs with emission at 365 nm and 395 nm driven by a current driver (MIC4801, Micrel Inc., San Jose, CA, USA) to adjust the excitation intensity. The fluorescence signal was collected by a miniature spectrometer (STS-Vis, Ocean Optics, Dunedin, FL, USA) with a 1024-element CCD array which acquires the entire spectrum in one single measurement with a resolution of 16 nm. The spectrometer was placed at 90° with respect to the LEDs to avoid the excitation light transmitted by the sample to reach the spectrometer.
The sensor has a recess where standard 4 ml clear glass vials with the sample can be inserted. 

All spectra of the dataset were acquired on undiluted samples at room temperature under identical conditions (illumination intensity, integration time, and geometry) for a quantitative comparison. The integration time was 1 s. 
During the measurements, the setup was kept in complete darkness to minimize the effect of stray light.

Each spectrum consists of an array of 1024 values (one for each pixel). The value corresponds to the intensity in counts at the different positions of the pixels. To obtain the wavelength (in nanometers) corresponding to each pixel, the following formula can be used:
\begin{equation}
    i = a + b\cdot i + c\cdot i^2 + d\cdot i^3
\end{equation}
where $i$ indicates the pixel ($i=0,...,1023$) and 
\begin{equation}
    \begin{split}
    a & = 337.92288208 \textrm{ nm} \\
    b & = 0.4470772743 \textrm{ nm} \\
    c & = 3.55128\cdot10^{-5} \textrm{ nm} \\
    d & = -8.38601\cdot10^{-9}\textrm{ nm} \\
    \end{split}
\end{equation}
Calibration parameters were provided by the spectrometer manufacturer.
All spectra correspond to the raw data without any data processing (smoothing, background subtraction, or normalization). Since all the measurements were done under identical conditions the intensities are directly comparable.

\subsection{Chemical Analysis}
For each olive oil sample, the dataset includes the values of the following chemical parameters: acidity, peroxide value, $K_{270}$, $K_{232}$, ethyl esters concentration and the samples quality class (EVOO, VOO, or LOO) (see Tab. \ref{tab:oils}). 

The chemical parameters were determined by accredited laboratories using the procedures described in the European Commission regulation and its amendment  (\cite{european2013commission,regulation1991commission}).



\vspace{6pt} 



\section{Funding}
This research was supported by the projects: “VIRTUOUS” funded by the European Union’s Horizon 2020 Project H2020-MSCA-RISE-2019 Grant No. 872181; “SUSTAINABLE” funded by the European Union’s Horizon 2020 Project H2020-MSCA-RISE-2020 Grant No. 101007702; “Project of Excellence” from Junta de Andalucia-FEDER-Fondo de Desarrollo Europeo 2018. Ref. P18–H0-4700.

\section{Author Contributions}
Conceptualization: Francesca Venturini and Umberto Michelucci; methodology: Francesca Venturini and Umberto Michelucci; software, Michela Sperti and Arnaud Gucciardi; validation, Francesca Venturini and Umberto Michelucci; formal analysis, Francesca Venturini and Umberto Michelucci; investigation, Francesca Venturini and Umberto Michelucci; resources, Vanessa M. Martos; data curation, Michela Sperti and Arnaud Gucciardi; writing, original draft preparation, Francesca Venturini and Umberto Michelucci; writing, review and editing, Francesca Venturini, Umberto Michelucci, Arnaud Gucciardi and Marco A. Deriu; funding acquisition, Vanessa M. Martos and Marco A. Deriu. All authors have read and agreed to the published version of the manuscript.

\section{Data Availability}

The data presented in this study are openly available in Dataset of Fluorescence Spectra and Chemical Parameters of Olive Oils at \url{https://data.mendeley.com/datasets/thkcz3h6n6/6}, DOI: 10.17632/thkcz3h6n6.6.

\section{Ackowledgments}
The authors would like to thank Michael Baumgartner and Ivo Herzig (Institute of Applied Mathematics and Physics, Zurich University of Applied Sciences, Winterthur, Switzerland) for help for the realization of the sensor, and Josep Palau Caballero and Arturo Jimenez (SCA San Sebastián Puente del Ventorro, s/n, 18566 Benalua de las Villas, Spain) for providing the oil samples.

\section{Conflicts of Interest}
The authors declare no conflicts of interest and no known competing financial interests or personal relationships that could have appeared to influence the work reported in this paper.



\section{Abbreviations}
The following abbreviations are used in this manuscript:\\

\noindent 
\begin{tabular}{@{}ll}
LOO & Lampante Olive Oil\\
EVOO & Extra Vigrin Olive Oil\\
VOO & Virgin Olive Oil\\
CCD & Charge-Coupled Device\\
LED & Light Emitting Diode\\
UV & Ultraviolet\\
FAEES & Fatty Acid Ethyl Ester\\
\end{tabular}


\bibliographystyle{unsrtnat} 
\bibliography{references}  

\begin{thebibliography}{4}
\providecommand{\natexlab}[1]{#1}
\providecommand{\url}[1]{\texttt{#1}}
\expandafter\ifx\csname urlstyle\endcsname\relax
  \providecommand{\doi}[1]{doi: #1}\else
  \providecommand{\doi}{doi: \begingroup \urlstyle{rm}\Url}\fi

\bibitem[Commission(2013)]{european2013commission}
European Commission.
\newblock Commission implementing regulation no 1348/2013 of december 17 2013.
\newblock \emph{Official Journal of the European Union}, 338:\penalty0 31--67,
  2013.

\bibitem[Commission(1991)]{regulation1991commission}
European Commission.
\newblock Commission regulation (eec) no. 2568/91 of 11 july 1991 on the
  characteristics of olive oil and olive-residue oil and on the relevant
  methods of analysis official journal l 248, 5 september 1991.
\newblock \emph{Offic. JL}, 248:\penalty0 1--83, 1991.

\bibitem[Venturini et~al.(2023)Venturini, Sperti, Michelucci, Gucciardi,
  Martos, and Deriu]{venturini2023extraction}
Francesca Venturini, Michela Sperti, Umberto Michelucci, Arnaud Gucciardi,
  Vanessa~M Martos, and Marco~A Deriu.
\newblock Extraction of physicochemical properties from the fluorescence
  spectrum with 1d convolutional neural networks: Application to olive oil.
\newblock \emph{Journal of Food Engineering}, 336:\penalty0 111198, 2023.

\bibitem[Venturini et~al.(2021)Venturini, Sperti, Michelucci, Herzig,
  Baumgartner, Caballero, Jimenez, and Deriu]{venturini2021exploration}
Francesca Venturini, Michela Sperti, Umberto Michelucci, Ivo Herzig, Michael
  Baumgartner, Josep~Palau Caballero, Arturo Jimenez, and Marco~Agostino Deriu.
\newblock Exploration of spanish olive oil quality with a miniaturized low-cost
  fluorescence sensor and machine learning techniques.
\newblock \emph{Foods}, 10\penalty0 (5):\penalty0 1010, 2021.

\end{thebibliography}






\end{document}